\documentclass[twocolumn]{revtex4-1}

\usepackage{amssymb}
\usepackage{amsmath}
\usepackage[T1]{fontenc}
\usepackage[utf8]{inputenc}

\newcommand{\pa}{\partial}

\newcommand{\myref}[1]{(\ref{#1})}

\newcommand{\om}{\omega} 
\newcommand{\Om}{\Omega}
\newcommand{\de}{\delta}
\newcommand{\De}{\Delta}
\newcommand{\al}{\alpha}
\newcommand{\eps}{\varepsilon}
\newcommand{\la}{\lambda}
\newcommand{\La}{\Lambda}
\newcommand{\ga}{\gamma}

\newcommand{\te}{\theta}

\renewcommand{\geq}{\geqslant}

\renewcommand{\hat}[1]{\widehat{#1}}
\newcommand{\lan}{\langle}
\newcommand{\ran}{\rangle}

\newcommand{\demi}{\frac{1}{2}}

\newcommand{\mcal}[1]{\mathcal{#1}}

\newcommand{\puiss}[1]{$^{#1}$}

\newlength{\somme}
\newlength{\sommep}
\newcommand{\intvide}{\rule[-\sommep]{0cm}{\somme}}

\newlength{\sommebis}
\newlength{\sommepbis}
\usepackage{empheq,fancybox}
\usepackage{color}
\usepackage[T1]{fontenc}
\usepackage{amsmath}
\usepackage{amssymb}
\usepackage{anysize}
\usepackage{graphicx}
\usepackage{bm}

\usepackage{ulem}

\usepackage{empheq}
\usepackage[most]{tcolorbox}

\newtcbox{\mymath}[1][]{%
    nobeforeafter, math upper, tcbox raise base,
    enhanced, colframe=blue!30!black,
    colback=blue!30, boxrule=1pt,
    #1}

\marginsize{2cm}{2cm}{2cm}{2cm}

\begin{document}
\title{Counter-rotation of magnetic beads in spinning fields}


\author{Jean Farago}
\author{Thierry Charitat}%
\author{Alexandre Bigot}
\author{Romain Schotter}
\author{Igor Kuli\'c}
\email{Also at Leibniz Institute for Polymer Research (IPF), 01069 Dresden, Germany.}
\affiliation{%
  Université de Strasbourg, CNRS, Institut Charles Sadron UPR-22, Strasbourg, France}


\date{\today}
\begin{abstract}
A magnetic stirrer,  an omnipresent device in the laboratory, generates a spinning magnetic dipole-like field that drives in a contactless manner the rotation of a ferromagnetic bead on top of it. We investigate here the surprisingly complex dynamics displayed by the spinning magnetic bead emerging from its dissipatively driven, coupled translation and rotation. A particularly stunning and counter-intuitive phenomenon is the sudden inversion of the bead's rotational direction, from co- to counter-rotation, acting seemingly against the driving field, when the stirrer's frequency surpasses a critical value.
 The bead counter-rotation effect, experimentally described in [{\it J.Magn.Magn.Matter}, {\bf 476}, 376-381, (2019)], is here comprehensively studied, with numerical simulations and a theoretical approach complementing experimental observations. 
\end{abstract}
\maketitle

\section{Introduction}

The broad availability of magnetic neodymium beads has led to an increased "table top" experimental interest in their self-assembly and individual dynamics \cite{vella2014magneto, Messina_2015,Becu2017,egri2018self}. Merely rolling such a bead on the top of the table leads to surprisingly complex behavior, with bead trajectories that depend on the motion speed and inclination to the local earth magnetic field \cite{unpublished}. 
In the microscopic realm, the interaction of ferro- and superparamagnetic beads with external fields and surface confinement has been extensively investigated. 
Spatially inhomogeneous, dynamic magnetic fields that are linearly propagating \cite{Tierno} or rotationally spinning  \cite{Gissinger,Becu2017} along surfaces have been investigated. The confinement of magnetic or magnetizable objects to solid \cite{Tierno,Becu2017} and fluid interfaces \cite{Snezhko,grzybowski2000dynamic} under oscillating fields leads to an intricate phenomenology including self-assembly and self-propulsion \cite{MartinSneszkoREVIEW}.

A permanently magnetized object placed in a non-uniform field, moves to minimize its magnetic free energy via two mechanisms: a) By aligning its magnetic moment with the field direction and b) by moving towards the maximum of field intensity. However, when coupled to a substrate the two otherwise independent degrees of freedom become tightly coupled giving rise to subtle and often counterintuitive effects.
Here we investigate one such easily reproducible, yet stunning effect : A neodymium bead placed on top of a laboratory magnetic stirrer. When the stirrer's magnet rotates at slow rates the bead naturally follows the field. However, surprisingly, once the field rotates fast enough the bead inverts its direction and rolls, to the surprise of the observer, in the opposite direction - against the driving field direction.  This effect was  recently described by Chau et al. \cite{Chau} and in a related form by Gissinger \cite{Gissinger}.  

In this paper, we revisit the experiments of Chau et al. \cite{Chau} with a comprehensive approach which combines experiments, numerical simulations and theoretical analysis. The benefit of numerical simulations is to provide  a complete description of the rotation of the beads, which are quite complicated to access experimentally. We performed  experiments, rather similar to the ``opposite polarity case'' described in \cite{Chau}, but simpler as we used a usual magnetic stirrer (commonly found in chemical laboratories) to provide the rotating magnetic field. We then wrote down the dynamical equations of a magnetic bead moving on the horizontal plane of the stirrer, assuming a viscous friction between the bead and the stirrer. The features of the magnetic field  of the stirrer have been carefully modeled, in order to reproduce as faithfully as possible the experiments. We simulated the dynamical equations obtained and by fitting two parameters of the model, we were able to reproduce qualitatively and quantitatively our experimental results. A theoretical analysis of the motion allows us (i) to understand the radial stabilization of the bead in its counter-rotating motion, (ii) to confirm the asymptotic role of the free rolling at large driving angular velocities, and (iii) to show that counter-rotating motion is not possible when purely paramagnetic beads (without remanent magnetization) are used.

The paper is organized as follows. In  sections \ref{descr} and \ref{expr}, the  system is introduced and our experimental results are presented. The dynamical equations of the system and the modeling of the rotating magnetic field are derived in section \ref{modell}. Two typical motions are then discussed in detail in section \ref{analysisofmotion}. A theoretical discussion follows in section \ref{theo}. For the sake of completeness, equations for the rotational dynamics of the bead in spherical coordinates are reported in Appendix \ref{app1}.

\begin{figure}[h]
  \centering
  \includegraphics[width=0.5\textwidth]{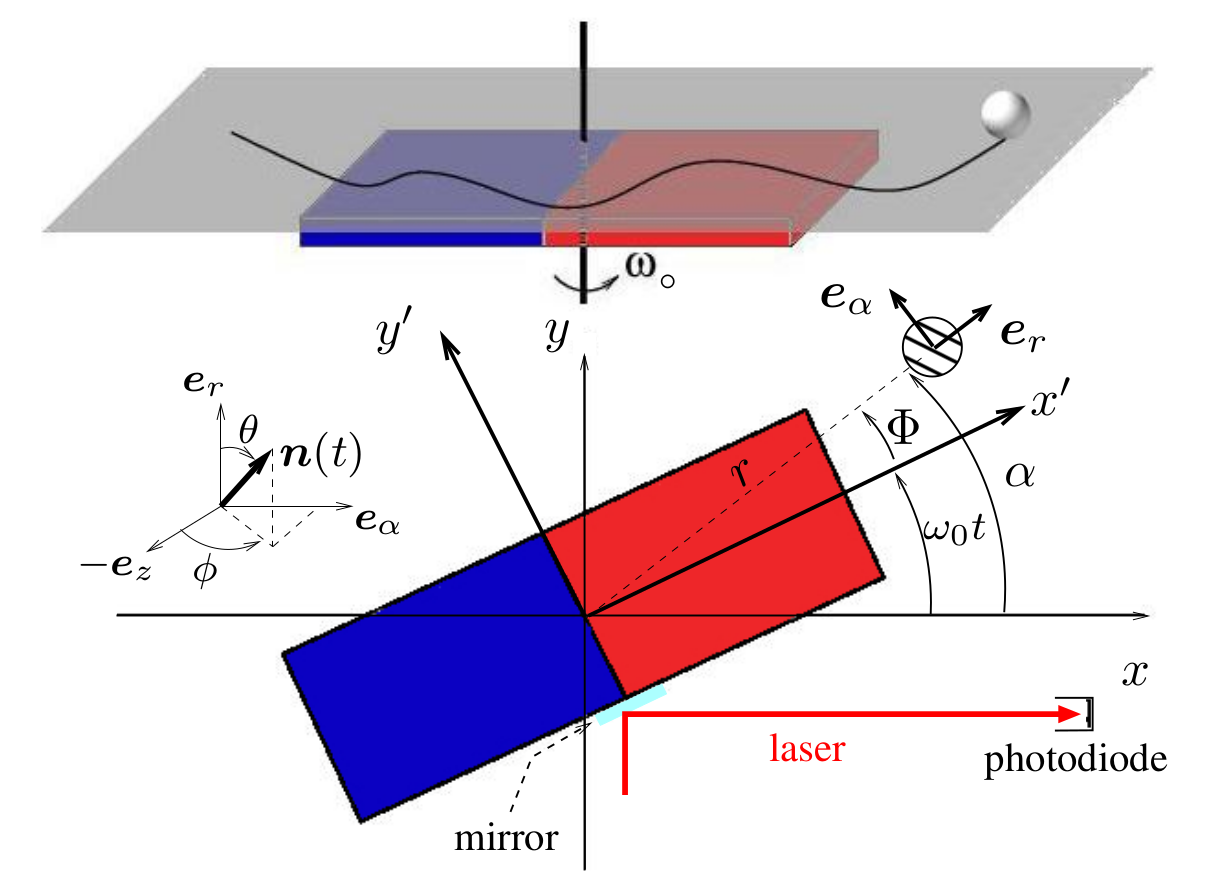}
  \caption{Sketch of the system and the main notations used throughout the paper. $\theta$ and $\phi$ are used to orient the magnetic axis $\bm n$ of the bead : $\bm n=\cos(\te)\bm e_r+\sin(\te)(\sin(\phi)\bm e_\al-\cos(\phi)\bm e_z)$.}
  \label{fig:0}
\end{figure}

\section{Description of the system}\label{descr}

The system, depicted in fig. \ref{fig:0}, is  a ferro\-magnetic (neodymium) sphere of mass $m=0.5$ g, radius $R=2.5$ mm and  magnetic moment along a diameter $\bm \mu(t)=\mu \bm n(t)$ (with $\bm n(t)$ a unit vector going from the south pole of the magnet to the north pole), placed on the stirrer surface (substrate). The latter is immobile in the laboratory frame $(O,\bm e_x,\bm e_y,\bm e_z)$ and located at the height $z=0$. The position of the sphere is described by a two-dimensional vector $\bm r_0=x_0\bm e_x+y_0\bm e_y=r_0 \bm e_r$, its vertical coordinate  staying at $z_0=R$. In the following, the local polar vector basis is denoted by ($\bm e_r,\bm e_\al$) and the corresponding cylindrical coordinates by $(r_0,\al_0,z_0)$.

Below the substrate, at the coordinate $z_0=-h_m=-16.2$ mm, a permanent magnet, of approximately rectangular shape (with a length $2\ell_m=56$ mm, a width $w_m=40$ mm, and thickness $9$ mm), rotates counterclockwise around the axis $Oz$ with a constant angular speed $\om_0$. Its time-dependent magnetic field $\bm{\mcal{B}}$ influences the spherical bead via the interaction potential $V(\bm n,\bm r_0,t)=-\mu \bm n\cdot\bm{\mcal{B}}(x_0,y_0,R,t)$. 
Notice that this interaction potential is an approximation which assumes that the magnetic field can be considered constant over the volume of the sphere.

\begin{figure}[h]
  \centering
  \includegraphics[width=0.5\textwidth]{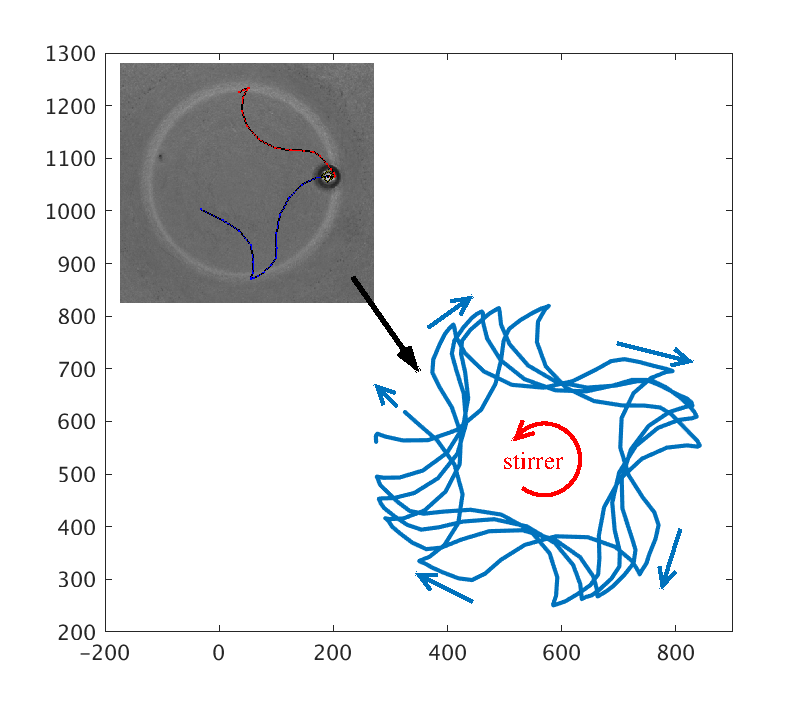}
  \caption{Example of a counter-rotating trajectory tracked by Blender (coordinates in pixels). The frequency of the rotating magnet is $\om_0/2\pi=10.7$ Hz.}
  \label{fig:blender}
\end{figure}
\section{Experimental results}\label{expr}

We recorded the bead trajectories using a high-speed camera (Phantom Miro LC320S) at 200 fps, and extracted them using the tracking features of the open-source software {\em Blender}, see fig. \ref{fig:blender} \cite{blender_software}. We used the reflection of a laser beam on a mirror glued on  the rotating magnet (cf. fig. \ref{fig:0}) to measure precisely the rotating frequency $\om_0$.

At very low rotation frequencies, the bead is trapped above one of the two poles of the rotating magnet (where the accessible magnetic field is maximum), and the bead's motion is trivially a corotation at the same angular speed as the rotating magnet. The proper rotation of the sphere is largely dictated by the requirement that the bead axis stays parallel to the local magnetic field. This proper motion yields a substantial friction between the bead and the horizontal slab of the stirrer. Beyond a certain angular velocity $\om_c$, the friction force is too high and the bead is no longer trapped in the magnetic potential. The typical value of $v_c$ is obtained by balancing the friction force $m\ga R\om_c$ with the magnetic force at stake $\mu B/\ell$. We get $\om_c\simeq \mu B/[m\ga R\ell]$. Fed with typical numerical values of our experiment (see values given above and in section \ref{NSR}), we obtain $\om_c/(2\pi)\simeq 7$ Hz, which is in accordance with what we observed in our experiment. For frequencies slightly above this limiting value, the behaviour is complex, mainly chaotic. This window is however rather narrow, and as can be seen in fig. \ref{fig:blender} for $\om_c/(2\pi)\simeq 10$ Hz, a regular regime (with some precession) sets in, where the global motion of the bead is counter-rotating, with a pattern , whose amplitude is large at small frequencies and shrinks at higher frequencies (see fig. \ref{fig:om04}, which corresponds to $\om_0/(2\pi)=19.9$ Hz. We term these patterns ``festoons'' in the following because of their similarity with the garland-like adornment of some architectural friezes \footnote{As remarked in \cite{Chau}, these festoons are locally similar to the family of mathematical curves named trochoids, or more precisely to hypocycloids, which  are constructed as the locus of a point of a disk rolling inside and against another, larger one.}.

 We measured as a function of the magnet rotation angular frequency $\om_0$  (i) the mean radius $r_0$ of the trajectories (ii) its standard deviation $\de r_0=\sqrt{\lan r_0^2\ran-\lan r_0\ran^2}$ and (iii) the mean angular velocity $\dot\al_0$ of the bead.
 
\begin{figure}[h]
  \centering
  \includegraphics[width=0.45\textwidth]{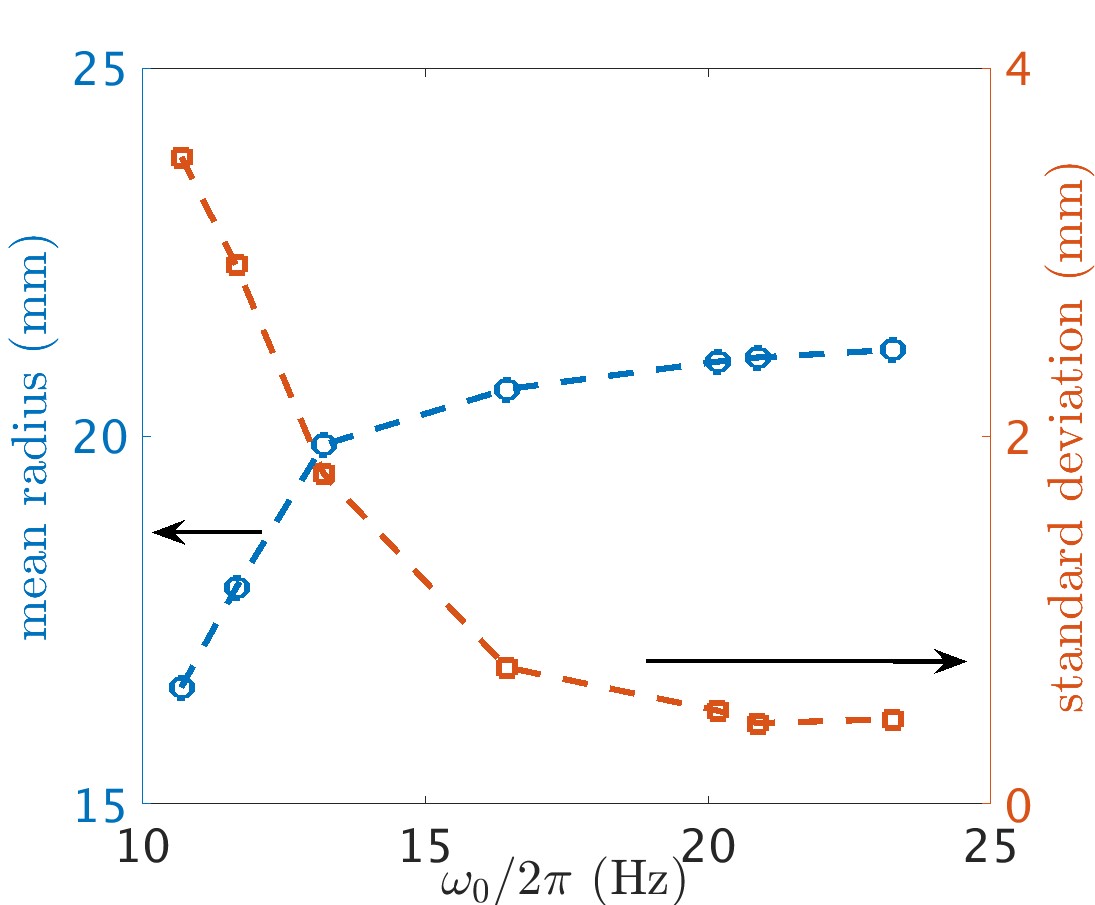}
  \caption{Blue circles : Mean radius $\lan r_0\ran$ of the trajectory (left plot). Red squares : Standard deviation $\sqrt{\lan r_0^2\ran-\lan r_0\ran^2}$ (right plot). Both curves are plotted as a function of the magnet frequency $\om_0/2\pi$. Only the counter-rotative regime is shown. The dashed lines are a guide for the eye.}
  \label{fig:radius}
\end{figure}
\begin{figure}[h]
  \centering
  \includegraphics[width=0.5\textwidth]{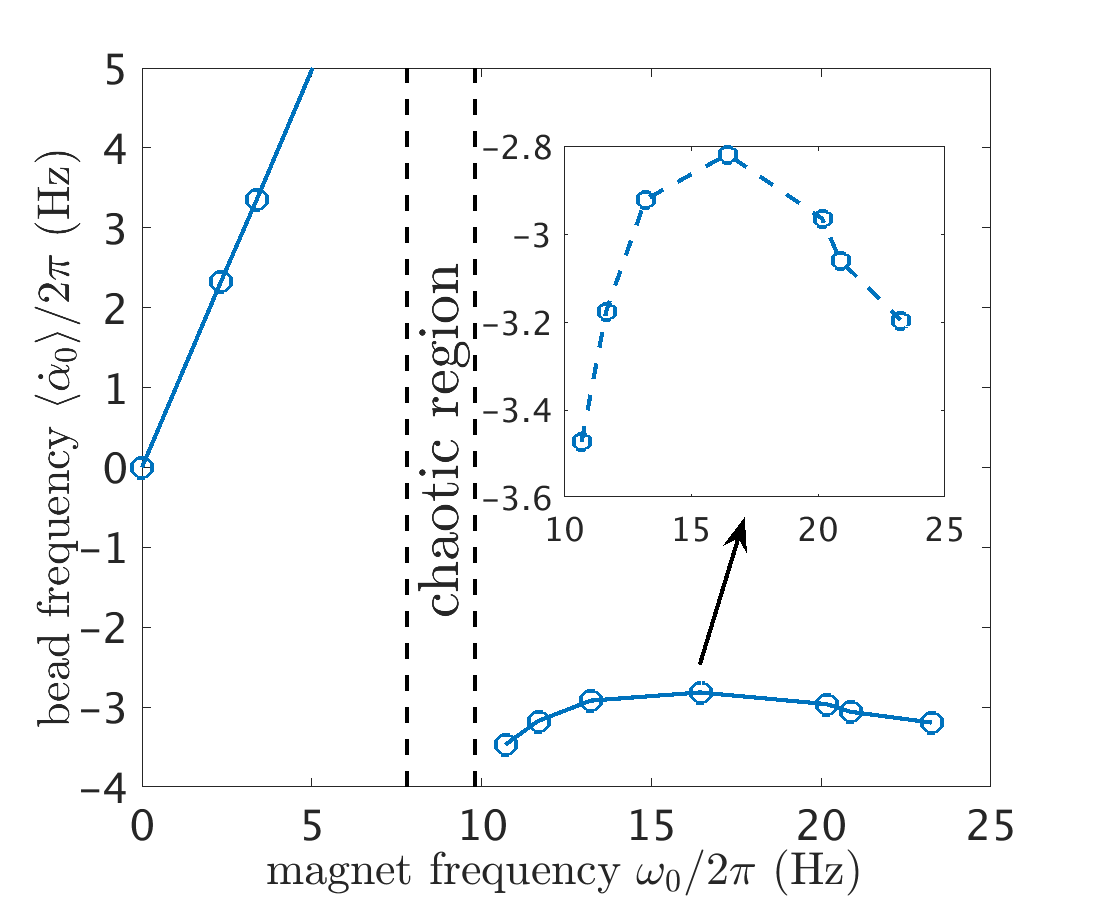}
  \caption{Mean bead frequency $\lan \dot\al_0\ran/2\pi$ as a  function of the magnet frequency $\om_0/2\pi$. Notice that $\lan\dot\al_0\ran$ is   negative in the regime of counterrotation. The lines are a guide for the eye and the inset is a zoom of the counterrotating region.}
  \label{fig:dotal}
\end{figure}
The mean radius and standard deviations are shown in fig. \ref{fig:radius}. The large standard deviation at small frequencies come from the large festooning of the trajectory, as can be seen in the example of fig. \ref{fig:blender}. When the frequencies become too small, the festoons cannot grow indefinitely and a chaotic behaviour is instead observed (for still lower frequencies, a co-rotative locked motion is recovered).
At larger frequencies, the festoons are still present, but with smaller amplitudes.

In fig. \ref{fig:dotal} we plot the  revolution frequency of the bead  $\lan \dot\al_0\ran/2\pi$ against the magnet frequency $\om_0/2\pi$. The corotative low frequency regime is rather obvious and described by $\dot\al_0=\om_0$, since it corresponds to the bead locked in one or another potential energy minimum, and constrained to follow the rotation of the magnet with quite a lot of frictional dissipation. At larger frequencies, after a chaotic transitional zone (where nothing relevant has to be reported, the bead being most of the time ejected from the stirrer table), the counterrotative region (characterized by negative $\dot{\al}_0$), the typical values of $|\dot\al_0|$ are one order of magnitude smaller than $\om_0$, which is  qualitatively explained by the mechanism which allows the counter-rotation : On the one hand, a frequency locking occurs between the rotation of the sphere around itself at an angular velocity, say $\lan\dot\phi\ran$, and $\om_0$ : $\lan\dot\phi\ran\sim \om_0$ (For this qualitative argument, there would be no need to define precisely $\dot\phi$, but a precise definition can be anyway given by looking at the definition of $\phi$ in fig. \ref{fig:0}). On the other hand, at large values of $\om_0$, the effect of the magnetic field averages rapidly to zero, so the motion must converge to a free rolling (albeit constrained into a circular motion) for which the friction on the table, proportional to the coincidental point velocity $\sim r_0\dot\al_0+R\dot\phi$, is approximately zero. As a result, we have $\dot\al_0\sim -\om_0R/\lan r_0\ran\sim -\om_0(R/\ell_0)$, for which in our case we have moreover $R/\ell_0\simeq 0.1$. The other salient feature of the right branch of fig. \ref{fig:dotal} is its bell shape with a maximum around $\om_0/2\pi=16$ Hz. This minimum signals a crossover between a complex regime  with few large festoons which act as shortcuts during revolutions (and therefore enhance the absolute value of the revolution frequencies), and a second regime ($\om_0/2\pi>16$ Hz) where the trajectories are close to circles, with many small festoons. For this regime, the previous arguments leading to $\dot\al_0\simeq -\om_0(R/\ell_0)$ apply and explain the enhancement of $|\dot\al_0|$.

\section{Modelling}\label{modell}

To achieve a comprehensive description of the bead motion in the counter-rotative regime, we develop a detailed theoretical model that we solve numerically.

Let us term $\bm \Om_0$ the rotation vector of the bead in the laboratory frame. We have for the time derivative in this frame $d\bm n/dt=\bm\Om_0\times\bm n$ which implies that $\bm\Om_0=\bm n\times(d\bm n/dt)+\dot\psi_0\bm n$ where $\dot\psi_0$ is the rotation velocity of the bead around its magnetic axis $\bm n$.

The location of the center of the sphere is given by the cylindrical coordinates $(r_0,\al_0,z_0=R)$, such that the velocity of its center of mass is $\bm v_G=\dot r_0 \bm u_r+r_0\dot\al_0 \bm u_\al$.
From the Koenig's theorem, we get a unconstrained Lagrangian
\begin{widetext}
\begin{equation}
  \mcal{L}_{\rm uncstr.}(\bm n,r_0,\al_0,\dot{\bm{n}},\dot{r}_0,\dot{\al}_0,\dot\psi_0,t)  =\frac{m}{2}[\dot{r}_0^2+r_0^2\dot\al^2_0]+\frac{1}{5}mR^2\left(\intvide [\frac{d\bm n}{dt}]^2+\dot\psi_0^2\right)+\mu\bm n\cdot\bm{\mcal{B}}(r_0,\al_0,t)\label{lagr}
  \end{equation}
\end{widetext}
where we used the expression $2mR^2/5$ for the inertia moment of the sphere with respect to one diameter. Notice that the constraint $\bm n^2=1$ affects the vector $\bm n$ for all times, so that the actual Lagrangian which describes the frictionless dynamics is $\mcal{L}=\mcal{L}_{\rm uncstr}-\demi\La(t)\bm n^2$, the function $\La(t)$ being the Lagrange multiplier associated to this constraint.

The dynamics of the sphere is affected by a possible friction of the sphere on the table (it is not constrained to  roll only). This friction is modelled by a force $\bm F_{\rm fr}$ proportional to the velocity $\bm{\mcal{V}_0}$ of the coincidental point $I$ (the point of the sphere in contact with the table at any instant) :
\begin{align}
  \bm F_{\rm fr}&=-m\ga \bm{\mcal{V}_0}=-m\ga\left(\bm v_G-R\bm \Om_0\times\bm e_z\right)
\end{align}
This friction can be incorporated in a Lagrangian description by means of the so-called Rayleigh function
\begin{align}
  \mcal{F}&=\frac{m}{2}\ga \bm{\mcal{V}_0}^2
\end{align}
which modifies the Lagrange equations to
\begin{align}
  \frac{d}{dt}\frac{\pa \mcal{L}}{\pa \dot q}&=\frac{\pa\mcal{L}}{\pa q}-\frac{\pa\mcal{F}}{\pa \dot q}
\end{align}
for all variables $q$ describing the dynamics.

 The lengths are made dimensionless by defining $\bm r =\bm r_0/\ell$ where $\ell$ is a characteristic length of the rotating magnet (we will choose $\ell$ slightly different from the actual length of the magnet $\ell_m$, as explained below). The magnetic field is also normalized according to $\bm B=\bm{\mcal{B}}/B_0$ ($B_0$ a characteristic magnetic field intensity of the rotating magnet). One defines $\eps=R/\ell$ and $\kappa=B_0\mu/[mR^2\ga^2]$. This last constant can be interpreted as the square of the characteristic time of friction  times the magnetic pulsation $\sqrt{\mu B_0/mR^2}$. A small value of $\kappa$ means that friction will likely overdamp the oscillations caused by $\bm B$ and its time evolution.  Finally the time is normalized by renaming $\ga t$ by $t$.  The outcome of these generalized dissipative and dimensionless Lagrange equations for $q\in\{\bm n,\dot\psi=\ga^{-1}\dot\psi_0,r,\al=\al_0\}$ is
\begin{align}
&  \bm\Om=\ga^{-1}\bm\Om_0=\bm n\times\frac{d\bm n}{dt}+\dot\psi\bm n,\label{Om}\\
&  \bm{\mcal{V}}=(\ell\ga)^{-1}\bm{\mcal{V}_0}=\dot{r}\bm e_r+r\dot\al\bm e_\al-\eps\bm \Om\times\bm e_z,\label{VV}\\
&  \ddot{\bm n}=-(\bm{\dot n})^2\bm n-\frac{5}{2\eps}(\bm{\mcal{V}}\times\bm e_z)\times \bm n+\frac{5\kappa}{2}[\bm B-\bm n(\bm n\cdot\bm B)],\label{nseconde}\\
 & \ddot{\psi}=-\frac{5}{2\eps}\bm{\mcal{V}}\cdot(\bm e_z\times\bm n),\\
  &\ddot{r}=r\dot{\al}^2+\kappa\eps^2\bm n\cdot\frac{\pa\bm B}{\pa r}-\bm{\mcal{V}}\cdot\bm e_r,\label{r}\\
  &\frac{d(r^2\dot\al)}{dt}=\kappa\eps^2\bm n\cdot\frac{\pa\bm B}{\pa \al}- r\bm{\mcal{V}}\cdot\bm{e}_\al.\label{sigz}
\end{align}
Notice that the derivative of $\bm{{B}}$ with respect to $\al$ includes the derivative of the unit vectors $(\bm e_r,\bm e_\al)$ as well as that of the coordinates $(B_r,B_\al,B_z)$ in case of a representation with cylindrical coordinates.

\begin{figure}[h]
  \centering
  \includegraphics[width=0.5\textwidth]{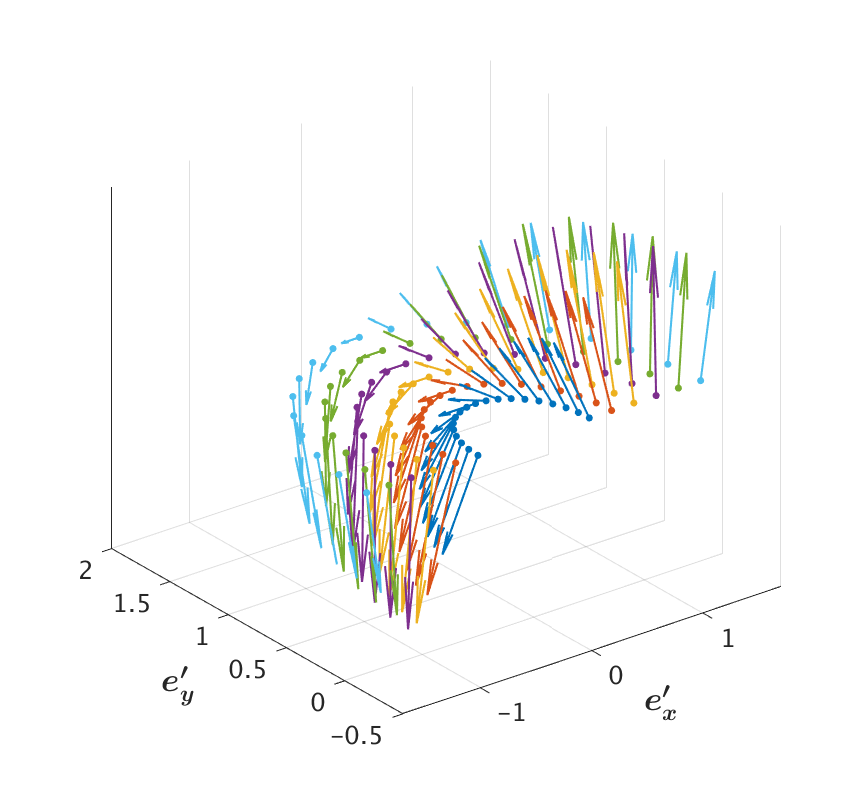}
  \caption{Magnetic field generated by five magnetic lines parallel to $\protect\overrightarrow{e_x'}$, of normalized length 1,  regularly placed at $y=n\la, n\in\{-2:2\}$, with $\la=w_m/(4\ell)=0.435$. }
  \label{fig:portrait}
\end{figure}

To model the magnetic field created by the rectangular rotating magnet, we assume it can be described by five parallel and equidistant lines of magnetic dipoles characterized by a constant  dipolar line-density $d\mcal{M}/dx'$ and a length $2\ell$. Their length $\ell$ is close to $\ell_m$, but is adjusted so that the two maxima of the magnetic field on the plate are separated by the same distance (41 mm) in the experiment and the modelling. We found $\ell=0.82\ell_m=23$ mm. On the other side, the distance between the two extremal lines are fixed to be exactly at $w_m$ - the width of the actual magnet. The actual portrait of the magnetic field experienced by the bead in shown in fig. \ref{fig:portrait}.  The field created at the  location $\bm r_0=(x_0,y_0,R)$ by a {\em single}  magnetic line of length $2\ell_0$, directed along the horizontal unit vector $\bm e_x'$, and symmetrical with respect to the point $(0,0,-h_w)$ is:
\begin{align}
  \bm{\mcal{B}}(\bm r_0,t)&= B_0\left[\frac{\bm r_0/\ell+(h_0/\ell)\bm e_z-s\bm e_{x'}(t)}{|\bm r_0/\ell+(h_0/\ell)\bm e_z-s\bm e_{x'}(t)|^3}\right]_{s=-1}^{s=+1}
\end{align}
where $ B_0=\frac{\mu_0}{4\pi\ell^2}\frac{d\mcal{M}}{dx'}$ and $h_0=h_w+R$. The normalized version $\bm B$ of this field is straightforwardly obtained by dividing by $B_0$ and writing the right hand side in terms of $\bm r=\bm r_0/\ell$ and $h=h_0/\ell$. The time dependence of the magnetic field is entirely borne by the vector $\bm e_x'$ which rotates counterclockwise in the laboratory frame $(O,\bm e_x,\bm e_y)$ : $\bm e_x'(t)=\cos(\om t)\bm e_x+\sin(\om t)\bm e_y$. Similarly the $(r,\al)$ dependence is given by the term $\bm r=r\bm e_r=r[\cos\Phi \bm e_x'+\sin\Phi\bm e_y']$ with $\Phi=\al-\om t$, see fig. \ref{fig:0}.

\subsection{Numerical simulations results}\label{NSR}

We simulated the dynamics of the bead embodied by equations (\ref{nseconde}-\ref{sigz}) using local spherical coordinates to represent $\bm n$, namely $\bm n= \cos\te\bm e_r+\sin\te(\sin\phi\bm e_\al-\cos\phi\bm e_z)$. The dynamical equations in these coordinates are given in appendix \ref{app1}. To compare quantitatively experiments and simulations, we have to fix the values of the parameters $\eps$, $\kappa$ and $\gamma$, the latter being involved in dimensionless angular velocities $\om=\om_0/\ga$ and $\dot\al=\dot\al_0/\ga$ for instance
. $\eps$ is an imposed geometric parameter : $\eps=R/\ell=0.11$. The other two are rather difficult to determine from experiments, all the more so the actual friction within the experiment is a solid friction, in contrast with our modeling of a linear viscous one. As a result, we chose to adjust $\kappa$ and $\ga$ so as to fit the experimental result at best. We found $\kappa=1.5$ and $\ga=312.5 \ s$\puiss{-1}. It gives a value $\mu B_0\sim 4.10^{-4}$ J. For a typical neodymium bead, we have $\mu\sim R^3\times BH_{\rm max}/ B_{\rm rem}$, where $BH_{\rm max}\sim 10^5$J$\cdot$m\puiss{-3} is the maximum energy product and $B_{\rm rem}\sim 1$ T the remanence. We get here $\mu\sim 10^{-2}$ A$\cdot$m\puiss{-1} and $B_0\sim 10^{-2}$ T.
\begin{figure}[h]
  \centering
  \includegraphics[width=0.5\textwidth]{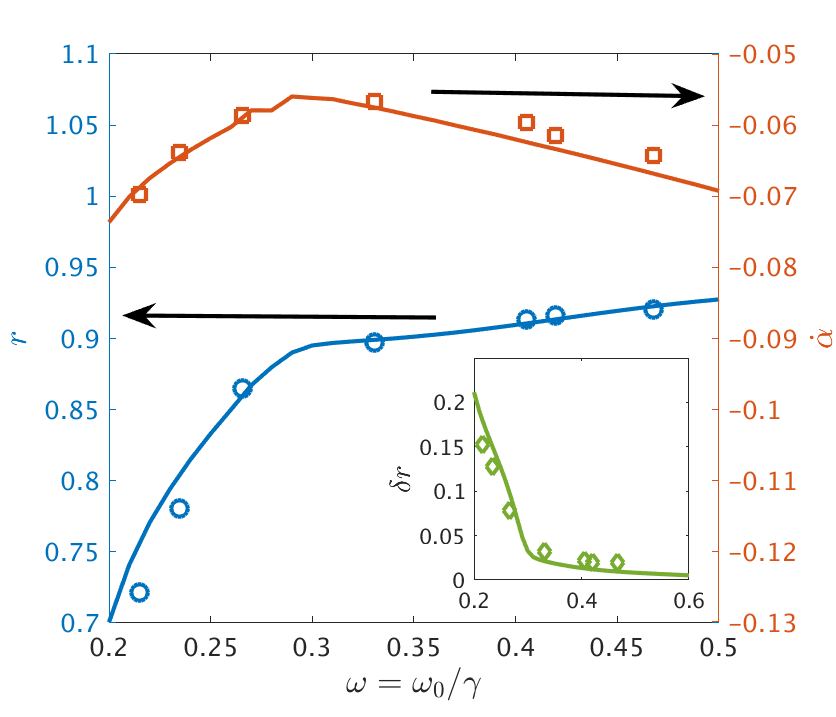}
   \caption{Comparison between experiments (symbols) and simulation (solid lines). Left ordinate and blue circles : Mean normalized radius $\lan r\ran$. Right ordinate and red squares : Mean normalized bead angular velocity $\lan \dot\al\ran$. Inset : Standard deviation $\de r=[\lan r^2\ran-\lan r\ran^2]^{1/2}$ vs. $\om$.}
   \label{fig:comparison}
\end{figure} 

With these values the agreement between experiment and simulations is quite quantitative, as can be seen in fig. \ref{fig:comparison}.

\section{Analysis of motion}\label{analysisofmotion}
\begin{figure}[h]
  \centering 
  \includegraphics[width=0.5\textwidth]{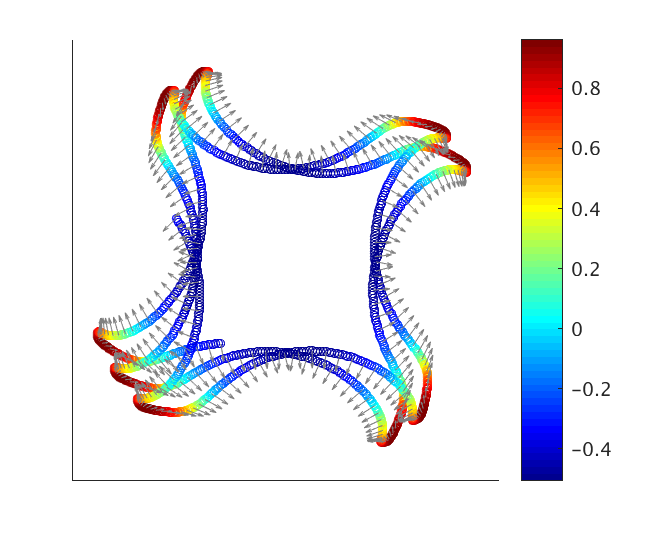}
   \caption{Simulation for the frequency $\om=0.215$ corresponding to the experiment of fig. \ref{fig:blender}. The color codes for the intensity of $n_z$ and the arrows show the direction and the relative length of the horizontal component of $\bm n$.}
   \label{fig:f4} 
  \end{figure}
 \subsection{4-fold counter-rotation}
The advantage of numerical simulations is to provide easily the  rotation of the bead. We first analyse the case represented in fig. \ref{fig:blender} corresponding to $\om=0.215$. In fig. \ref{fig:f4}, the simulated trajectory is represented with a color code corresponding to the values of $n_z\in[-1,1]$ (A movie of the simulation can be found in Supplemental Material [\verb!film_0215.avi!], where the motion of both the bead and the rotating magnet are shown). Small gray arrows are also added to see the direction and relative length of $\bm n-n_z \bm e_z$. It is  worth noting that the mean value of $n_z$ is not zero, which indicates that a different, conjugate  solution at this frequency exists where polarities of $\bm n$ and $\bm B$ are simultaneously reversed.  In this case the corners of the square-like shape of the trajectories would correspond to minima $n_z\simeq -1$.
\begin{figure}[h]
  \centering
  \includegraphics[width=0.5\textwidth]{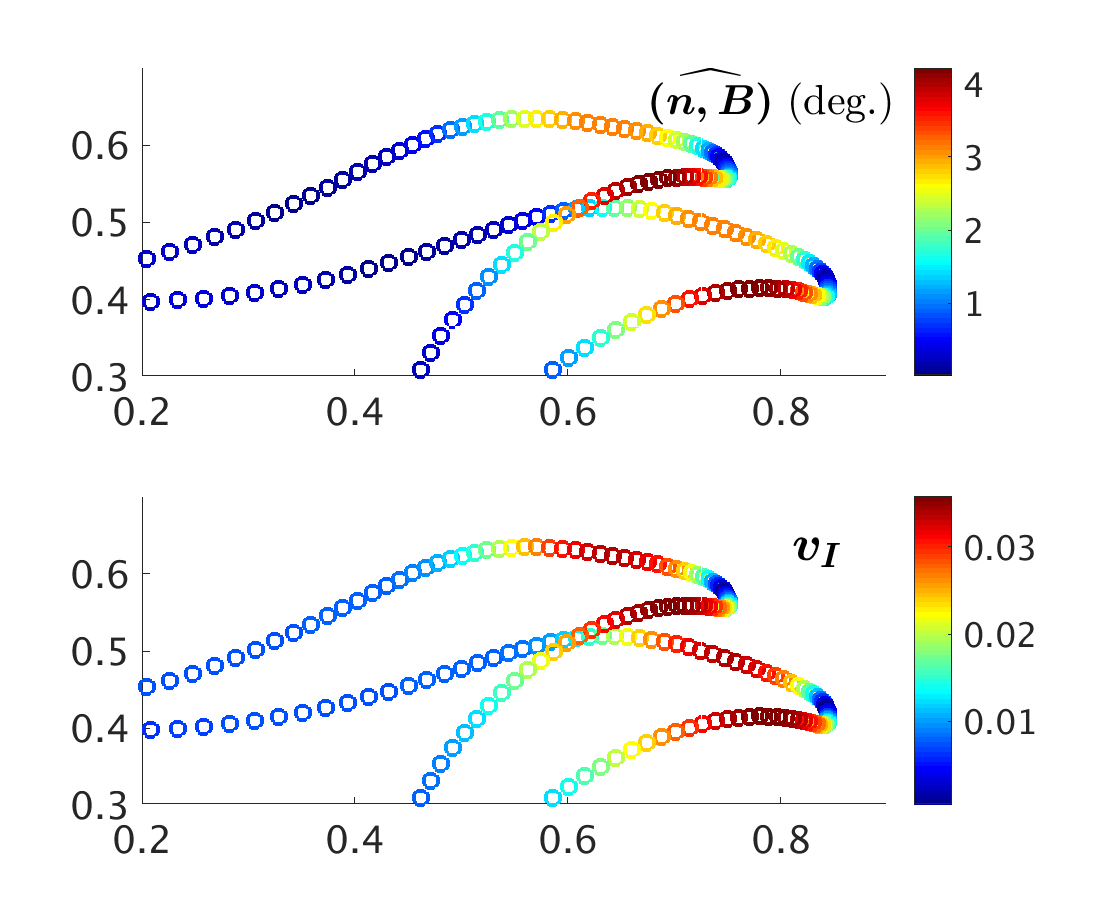}
   \caption{Same as fig. \ref{fig:f4} (upper right corner only), the color codes showing in the top plot the angle between the magnetic axis of the bead and the local magnetic field (in degrees), and in the bottom plot the magnitude of the coincidental point velocity, proportional to the friction force experienced by the bead.}
   \label{fig:f4detail}
\end{figure} 
It is worth mentioning that in this motion, the {bead's moment} stays remarkably parallel to the magnet field, as can be seen in the fig. \ref{fig:f4detail} (top) : {Their relative angle} does not exceed 4 degrees. The bottom plot shows the magnitude of the coincidental velocity. It can be seen that the high friction zones are tightly correlated to those (rare) moments where the bead axis cannot follow the rate of variation of the magnet's field. It is also interesting to note that the magnitude of the magnetic field experienced by the bead during its revolution does not vary more than 13\% with respect to its mean value (not shown).

{ Figure \ref{fig:f4detail} is interesting also because it allows a discussion on the linear viscous friction hypothesis. Why indeed does the model with a viscous friction describe so well the experiments ? We see from the figure \ref{fig:f4detail} (bottom) that the friction force acts on quite localized moments of the trajectory, close to the turning points. So one can expect that some dynamical details of the trajectories in the vicinity of the turning points may be only approximately accounted for (for instance, the turning points in the real $\om=0.215$ case (see for instance fig. \ref{fig:blender} where the turning points look sharper than those of fig. \ref{fig:f4}). On the other hand, the rest of the trajectories should be correctly described, provided the characteristic times associated to the energy dissipation of both friction mechanisms be comparable. For a dry friction with parameter $\mu_{\rm dyn}$, the characteristic time of dissipation is $\tau\sim I\dot\phi_0/(\mu_{\rm dyn} Rmg)$ where $\dot\phi_0\sim\om_0$. For a viscous dissipation, one has instead $\tau\sim I/[mR^2\ga]$. On equating both expressions, one finds $\ga\simeq \mu_{\rm dyn}g/[R\om_0]$. So, in principle $\ga$ depends on $\omega_0$, but in the  range $\om_0\simeq 10$ Hz where the festoons are well developed and therefore the friction is not negligible, it gives $\ga\simeq 10^2$ Hz, i.e. the  order of magnitude we got for $\ga$ by direct fit (we found $\ga=312.5$ s\puiss{-1}).

  }
  
\subsection{Many-fold counter-rotation}

\begin{figure}[h]
  \centering
  \includegraphics[width=0.5\textwidth]{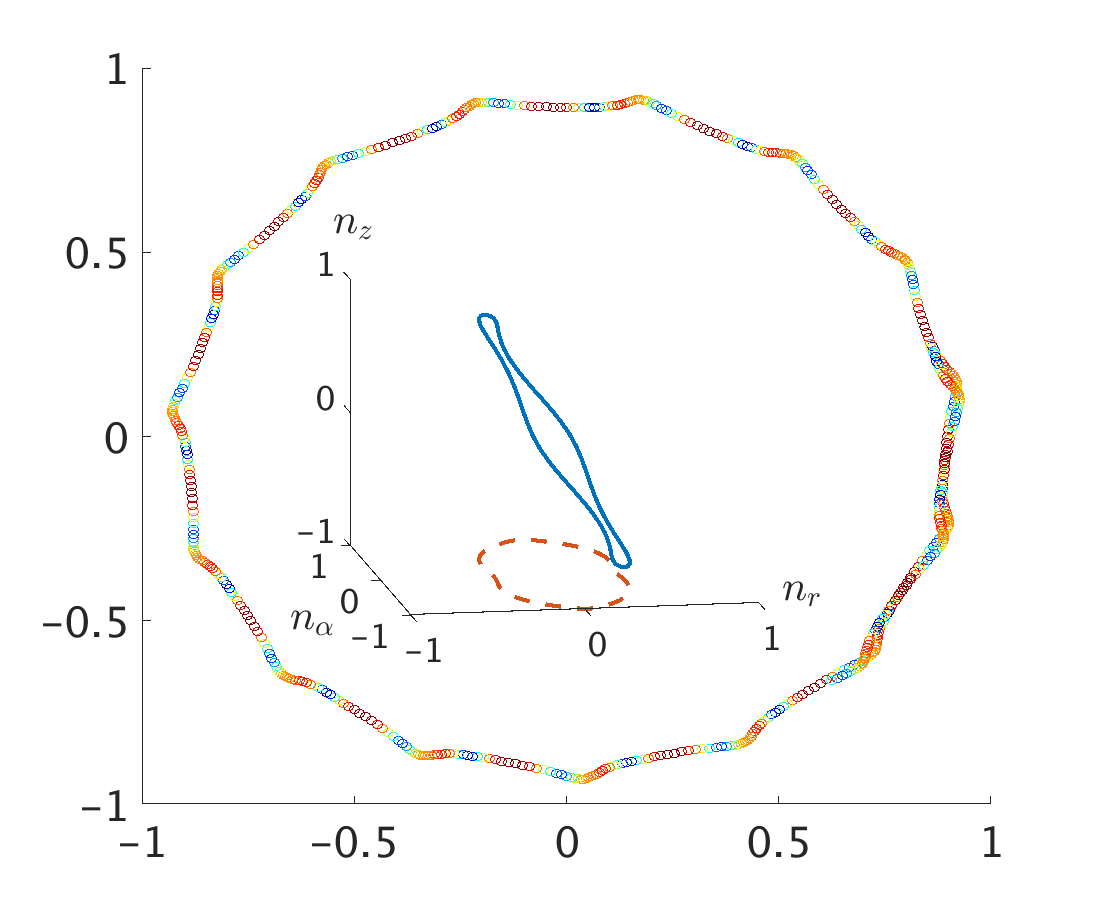}
   \caption{Trajectory of the magnetic bead for $\om=0.4$. The color modulates according to the magnitude of the friction force. The inset (solid blue line) shows the closed trajectory of the vector $\bm n$ in the local coordinate frame $(\bm{e}_r,\bm{e}_\al,\bm{e}_z)$.}
   \label{fig:om04}
\end{figure} 
For increasing values of $\om$, the counter-rotation tends to adopt a more circular shape, as can be seen from fig. \ref{fig:om04} where the trajectory of the bead for $\om=0.4$ is shown (a movie of the simulation can be found in Supplemental Material [\verb!film_04.avi!]). The festoons are numerous (15 for $\om=0.4$) and have a small amplitude. As before, the magnetic bead stays remarkably parallel to the local magnetic field (the angle is never larger than 3$^\circ$), and in the local frame $(\bm e_r,\bm e_\al,\bm e_z)$, the trajectory is closed and almost circular, with the magnetic moment vector displaying a mild tilt ($\simeq 25^\circ$) with respect to the vertical.

{
\subsection{Order of the patterns  and  period halving}

As can be seen in figs \ref{fig:f4} and \ref{fig:om04}, (i) the patterns have different orders corresponding to the number of festoons they are made of and (ii) in general, the patterns are slowly shifting and do not superpose when the bead has completed a revolution. 
The criterion to  have a truly periodic motion in the laboratory frame, with $N$ festoons reads $2\pi/(N|\dot{\al}_0|)=(2\pi/\om_0)(1-1/N)\Leftrightarrow N=a(1+ \om_0/|\dot{\al}_0|)$ with $a=1$ if the bead has a periodic motion in the rotating magnet frame requiring a full turn, i.e. $\bm r(\Phi+2\pi)=\bm r(\Phi)$ (symmetry S1). The possibility of a period halving exists if the trajectory has the finer symmetry $\bm r(\Phi+\pi)=-\bm r(\Phi)$ (symmetry S2, with the rotational reversal $\bm n(\Phi+\pi)=-\bm n(\Phi)$). In this case the number of festoons is related to the frequency by $N=a(1+\om_0/|\dot{\al}_0|)$ with $a=2$. By trial and errors we were able to find in the numerical simulations the frequencies $\om$ for which the pattern is approximately periodic in the laboratory frame. The results are shown in the table \ref{table1}.
\begin{table}[h]
  \centering
  \begin{tabular}[c]{|c|c|c|c|}
    \hline
    frequency $\om$ & \# of festoons & $a$ & $a(\om/|\dot \al| + 1)$ \\
    \hline
    0.21 & 4 & 1 & 3.98\\
    0.25 & 5 & 1 & 5.03\\
    0.2845 & 6 & 1 & 5.99\\
    0.352 & 14 & 2 &  13.98\\
    0.405 & 15 & 2 & 14.96\\
    \hline
  \end{tabular}  \label{table1}
  \caption{Laboratory frame periodic trajectory parameters}
\end{table}
One sees that on enhancing $\om$, a period halving transition occurs in the range $\om\in[0.2845,0.352]$, going from S1-invariant trajectories to S2-invariant ones.
\begin{figure}[h]
\centering  \includegraphics[width=0.5\textwidth]{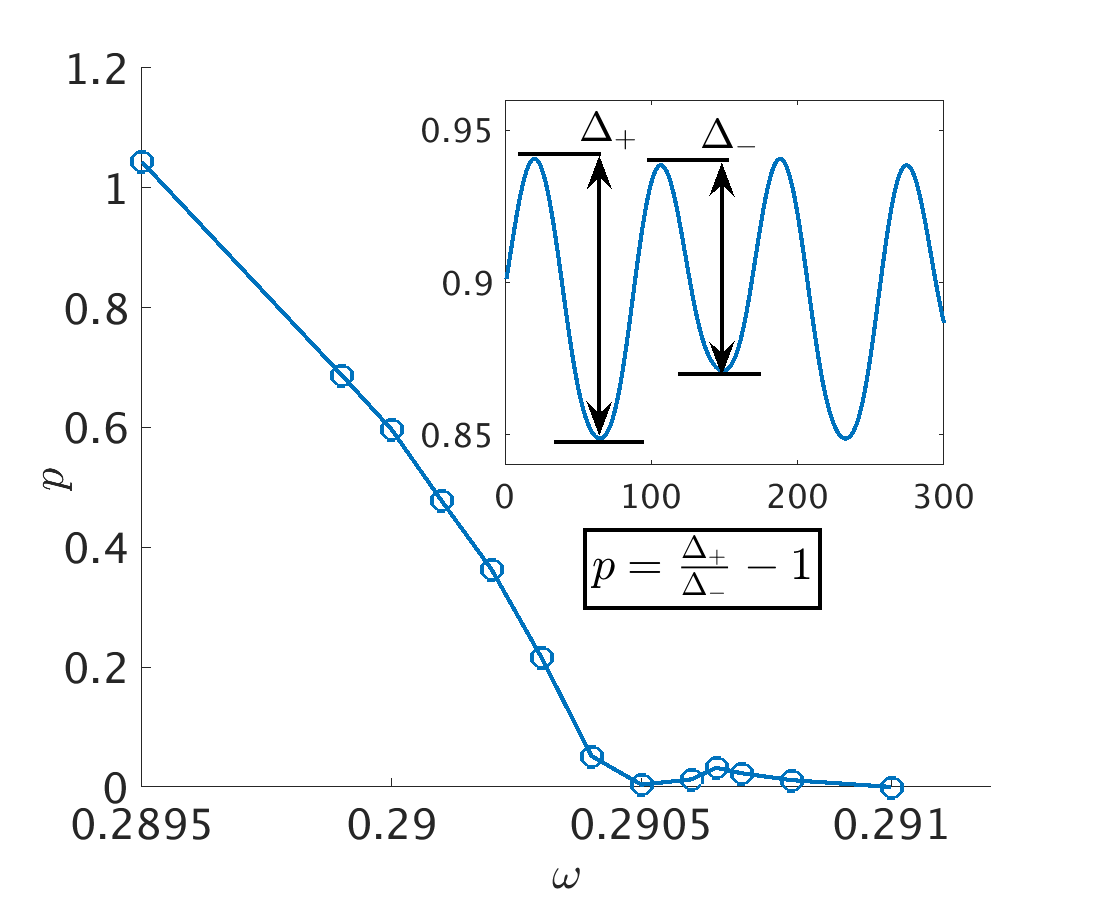}
  \caption{Supercritical transition of period halving near $\om=0.2904$. The inset shows $r(t)$ in the $p\neq 0$ region in order to highlight the definition of $p$ as the ordered ratio of two successive amplitudes in $r(t)$.}
  \label{supercritical}
\end{figure}
The precise location of the transition has been found for $\om\simeq 0.2904$ as can be seen in figure \ref{supercritical}, where the order parameter $p$ has been chosen as $p=\De_+/\De_--1$, namely the larger-than-one ratio of the amplitudes of two successive oscillations of the parameter $r$ (minus 1). One sees that the transition is supercritical, which is confirmed by the diverging relaxation time associated with the convergence of $p$ near the transition.

}

\section{Time averagings}\label{theo}

The full dynamical behaviour of the bead is complicated and certainly non integrable.

One can nevertheless try to make some predictions concerning the mean radius of counter-rotation and the associated rotation frequency, at least in the regime where $r$ stays reasonably constant.

The map of the field shows that close to the maximum, it has essentially an (slightly tilted) orthoradial structure. It is thus reasonable to assume that in the dynamical regimes where $r\simeq 1$, one can neglect the radial dependence of $\bm n$, assume $\te\simeq \pi/2$ and write $\bm n\simeq \sin\phi\bm e_\al-\cos\phi\bm e_z$. Likewise, we neglect also the fluctuations in $r$ and $\dot\al$. Writing again $r$ and $\dot\al$ for the temporal averages $\lan r\ran$ and $\lan \dot\al\ran$, we have
\begin{align}
  0&=r\dot\al^2+\kappa\eps^2\pa_r \left\lan B_\al\sin\phi -B_z\cos\phi\right\ran-\lan\mcal{V}_r\ran,\label{eq21}\\
  0&=\kappa\eps^2\left\lan [\pa_\al B_\al+B_r]\sin\phi-\pa_\al B_z\cos\phi\right\ran-r\lan\mcal{V}_\al\ran,\label{eq22}\\
     0&=\kappa\eps\lan B_\al\cos\phi+B_z\sin\phi\ran-\lan\mcal{V}_\al\ran.\label{eq23}
\end{align}
In the approximation considered, one has, from the formula \myref{Valpha} of the Appendix \ref{app1},   $\mcal{V}_\al\simeq r\dot\al+\eps\dot\phi$. 
\begin{figure}[h]
  \centering
  \includegraphics[width=0.5\textwidth]{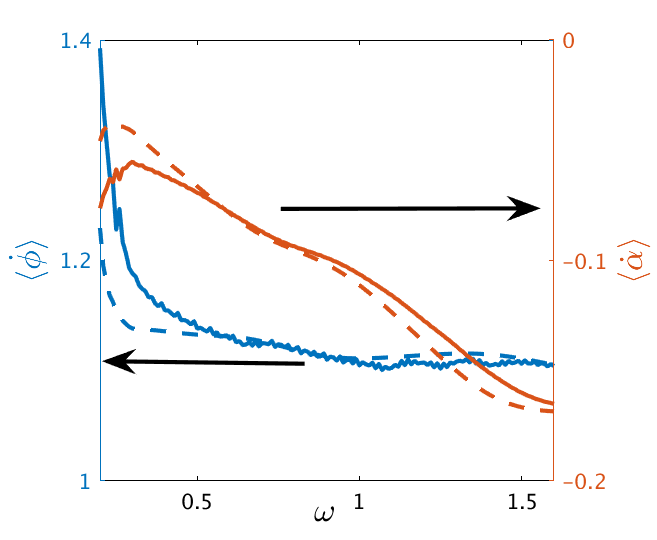}
  \caption{Test of the formulas \myref{predictions} : The solid lines show $\lan\dot\phi\ran$ (blue, left ordinate) and $\lan\dot\al\ran$ (red, right ordinate) and the dashed show the result of \myref{predictions}.}
  \label{fig:test}
\end{figure}
The equation \myref{eq21} shows that the centrifugal force is counterbalanced by a magnetic force  only if $\phi$ oscillates with the same frequency as $\Phi=\al-\om t$. This leads us to assume $\phi=-\Phi+\chi$ where $\chi$ is a constant phase. One can show (but the calculation is cumbersome) that the averages implying the magnetic field in eqs \myref{eq22} and \myref{eq23} are all $\propto \sin(\chi)$ for symmetry reasons, whereas that of \myref{eq21} is $\propto \cos(\chi)$. The solution of these equations is therefore somewhat simplified, since they reduce to (i) $\chi\equiv 0$ modulo $\pi$ and (ii) $\lan \mcal{V}_\al\ran=0$ and (iii) eq. \myref{eq21}.
Actually, one can guess in advance that the phase locks to $\chi=\pi$, because it corresponds to the most stable situation where the bead visits the region of maximum magnetic field in the orientation which minimizes the magnetic energy interaction. Combining $\dot\phi=\om-\dot\al$ and $r\dot\al+\eps\dot\phi=0$, we obtain
\begin{align}
 \dot\al=-\frac{\eps\om}{r-\eps}\text{ and }\dot\phi=\frac{\om r}{r-\eps}.\label{predictions}
\end{align}
The comparison of these formulas with the actual averages of $\dot\phi$ and $\dot\al$ are shown in fig. \ref{fig:test} and the result is convincing for $\om\geq 0.5$, that is for frequencies {\it higher} than those obtained in the experiments of fig. \ref{fig:comparison}. This means that the hypothesis of free rolling corresponding to the relation $r\dot\al+\eps\dot\phi=0$ is quantitatively correct only at quite large frequencies. Regarding the prediction for the mean value of $r$, one would use eq. \myref{eq21}, but this equation would be tractable only if $\lan\mcal{V}_r\ran$ is negligible with respect to the other terms, since the expression \myref{Vr} for $\mcal{V}_r$ contains a  term $-\eps\dot\psi\sin\te\sin\phi$ addressing directly the rotation of the bead around its magnetic axis, a motion that is coupled to all degrees of freedom.
\begin{figure}[h]
  \centering
  \includegraphics[width=0.5\textwidth]{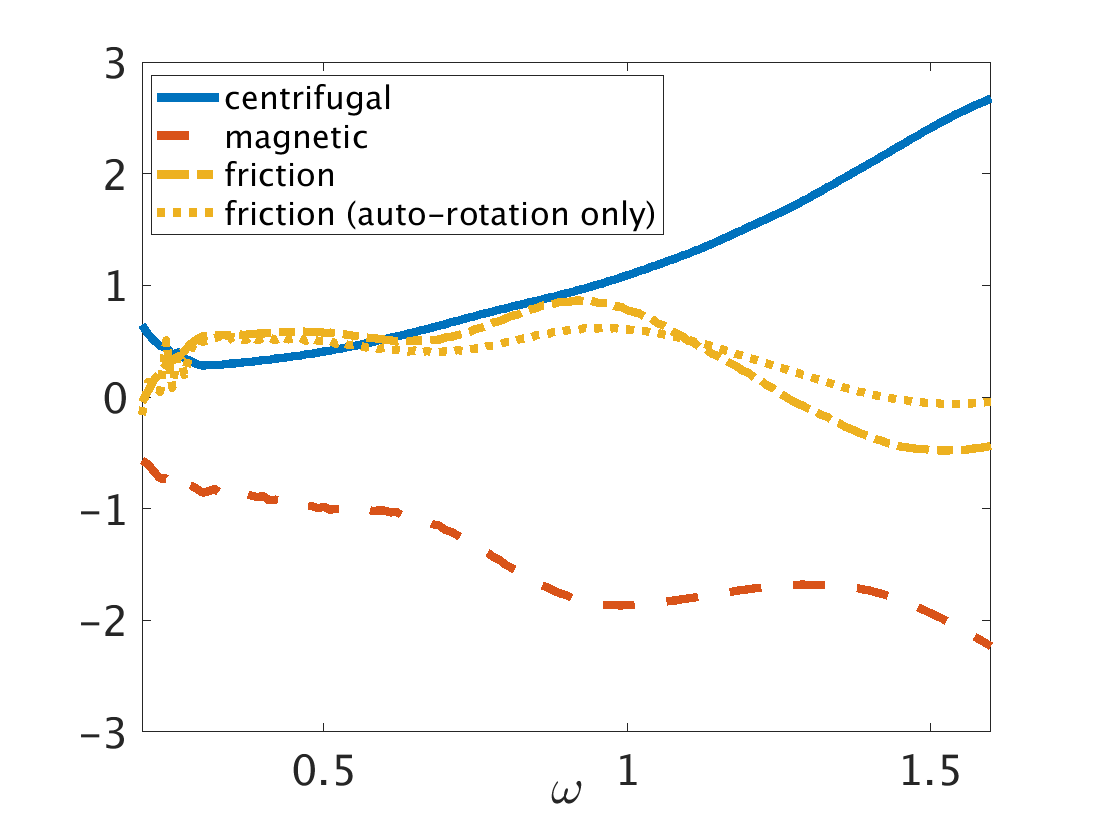}
  \caption{Evolution with $\om$ of the three terms of eq. \myref{eq21}. ``centrifugal'' refers to $r\dot\al^2$, ``magnetic'' to  $\kappa\eps^2\pa_r\lan B_\al\sin\phi-B_z\cos\phi\ran$ and ``friction'' to $-\lan\mcal{V}_r\ran$. The dotted yellow curve shows $\eps\lan\dot\psi\sin\te\sin\phi\ran$, the term of $-\lan\mcal{V}_r\ran$ depending on the rotation of the bead around its magnetic axis. All curves are divided by $\eps^2$.}
  \label{fig:evolution}
\end{figure}
As can be seen from the inspection of fig. \ref{fig:evolution}, the friction term $-\lan \mcal{V}_r\ran$ is not at all negligible in the regime $\om<0.5$ and becomes negligible with respect to the other two only at quite higher frequencies. As a result, one concludes that the quantitative features of the counter-rotating regime cannot be simply obtained in the moderate driving frequencies where the festooning of the trajectories is marked.

A final comment can be made about an implicit choice made in assuming $\phi=-\Phi+\chi$, a relation dictated by the requirement that $\phi$ and $\Phi$ must leads to resonant terms in the magnetic force. There is here an implicit because $\phi=\Phi+\chi$ would have been also a valid choice. In Appendix \ref{app:why} it is shown why this Ansatz, which would give a co-rotative regime, is actually never observed.

\subsection{Physical origin of the festoons}
\begin{figure}[h]
  \centering
  \includegraphics[width=0.4\textwidth]{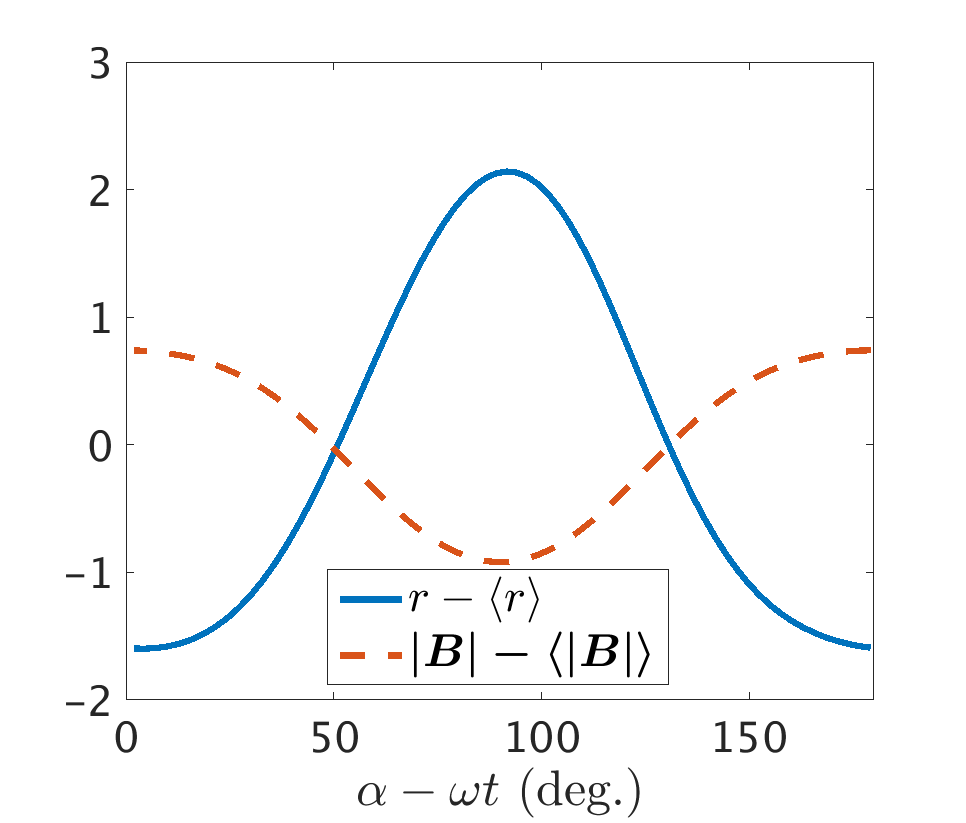}
  \caption{Fluctuations of $r$ (solid blue) and $|\bm{B}|$ (dashed red) during half a revolution of the magnet. The abscissa is the angle between the  bead and the rotating magnet. Notice that when the magnet and the bead are on top of each other, the value of $r$ is minimal and is $0.89$, i.e. the location of the maximum of the field. }
  \label{fig:magint}
\end{figure}
On the qualitative level, the origin of the festoons can be understood if one realizes that the magnetic axis of the bead stays always nearly colinear to the local magnetic field. As a result, the effective magnetic force for the bead's center of mass is high near the ends of the magnet where the field varies substantially over a short distance. As shown in fig. \ref{fig:magint}, one sees that the radius is minimal, around $0.89$ (the location of the absolute maximum of field), when the magnet crosses the bead angular position. When the magnet axis goes away from the bead angular position, the field variations weaken, the centrifugal force ``wins'' and drives the bead away from $r=0.89$, whence the maximum of $r$ at precisely $\al-\om t=\pi/2$. However, the detailed shape of the festoons cannot be accounted for by such a simple force balance argument, because in the vicinity of the maxima of $r$, the friction force is no longer negligible in the budget, as can be seen in fig. \ref{fig:f4detail} (bottom).

\subsection{Paramagnetic bead}


As correctly noticed in \cite{Chau}, the mechanism for the counter-rotation proposed by \cite{Gissinger} relies on the presence of a remanent magnetization in the beads, and the counter-rotation observed with steel beads would be entirely due to it. With the theory presented in this work and summarized by equations (\ref{Om}-\ref{sigz}), it is possible to test an ideal case where the magnetic interaction would be solely paramagnetic. It amounts to replacing the interaction potential in the Lagrangian by $V=-\al_{\rm m}B^2$. The most important consequence of this new interaction is that the rotational dynamics of the bead is now decoupled from the magnetic field by direct interaction, that is the term $\propto \kappa$ in \myref{nseconde} disappears. All the  arguments put forth previously to account for the counter-rotation are no longer valid, and we indeed never observed counter-rotation in simulations of the purely paramagnetic bead.

To have a theoretical indication (not a proof) of why the counter-rotating stationary is generally suppressed, we consider the time evolution of the energy function \cite{Goldstein} $h=\dot{q}\pa\mcal{L}/\pa\dot{q}-\mcal{L}$ is $dh/dt=-2\mcal{F}-\pa\mcal{L}/\pa t$ which yields after time averaging and noticing that $\pa_t B^2=-\om \pa_\al B^2$
\begin{align}
  \lan \mcal{V}_r^2+\mcal{V}_\al^2\ran=\al_m\om\left\lan\frac{\pa B^2}{\pa \al}\right\ran\label{tondu}
\end{align}
On the other hand, the time  averaging of eq. \myref{sigz} (with the first term of the right hand side replaced by $\al_m\pa_\al B^2$) yields $\al_m\lan\pa B^2/\pa \al\ran=\lan r\mcal{V}_\al\ran$. So that we have from \myref{tondu}
\begin{align}
  \om \lan r\mcal{V}_\al\ran >0\label{17}
\end{align}
Physically, it means that the friction force removes angular momentum from the particle with respect to the rotating magnet. Writing that $\lan d(\demi m \bm v_G^2)/dt\ran=0$, we have also
\begin{align}
  \om\al_m\lan \pa B^2/\pa \al\ran=\lan \dot{r}\mcal{V}_r+r\dot{\al}\mcal{V}_\al\ran>0
\end{align}
which shows that the work of the friction force on the sphere center of mass is resistive. in average. So, if one assumes, for high enough frequencies, a motion which is very close to a free rolling at a fixed distance from the magnet's center, it means that $r$ and $\dot{\al}$ are almost constant and that $\dot{r}\simeq 0$, whence the right hand side of the preceding equation is asymptotically $\sim \lan\dot\al\ran\lan r\mcal{V}_\al\ran$. If this is correct, we have both \myref{17} and $\lan\dot\al\ran\lan r\mcal{V}_\al\ran>0$ yielding $\lan\dot\al\ran \om>0$, i.e. the counter-rotation is impossible. Although not a mathematical proof, the argument is qualitatively correct, provided that the radial fluctuations are negligible in the high frequency regime, as well as those of $\dot\al$. It is worth noting however that the argument relies on the positivity of \myref{tondu}, which is of thermodynamical origin, since its corresponds to the dissipative work of the friction force. As a result, the argument should therefore apply to a pure paramagnetic bead experiencing a dry friction as well.

\section{Conclusion} 

In this paper, we have presented a coherent study of the {counter-rotation} of a ferromagnetic bead, constrained to move on a magnetic {stirrer's surface}, and excited by the rotation at constant angular velocity of the magnet installed beneath the slab of the stirrer. By fitting two parameters, we were able to  reproduce {quantitatively} by numerical simulations the experimental observations, first observed by \cite{Chau} in a very similar experiment,  {in spite of the different type of friction  of the bead on the slab utilized (viscous vs. dry friction).} The  expression of the dynamical equations of the ten degrees of freedom of the problem (two for the bead's center of mass, three for the orientation of the bead, plus the same number for their time derivatives), allowed us to make also some theoretical analysis in the regime of high frequencies. {We show in particular that the corotative regime is never stable at high frequencies (whereas it would be observable for a system of a magnetic disk holonomically constrained to roll at a fixed distance from the center) and furthermore that a purely paramagnetic and isotropic bead can never display counter-rotation}.  Therefore the slow counter-rotation observed by \cite{Chau} with steel {spheres are entirely attributable to the slight remanent magnetization or magnetic anisotropy of the spheres}. We further analyzed the dynamical behavior of the bead when festoons are present and showed that the associated modulation of the radial distance of the bead is tightly correlated to ---and therefore mainly due to--- the modulation of the radial component of the magnetic force : When this component weakens, the centrifugal force moves the bead away from the rotation center, and conversely.
This simple argument is asymptotically true only for $\om\rightarrow\infty$. At lower frequencies, the friction force is not negligible, cf. fig. \ref{fig:evolution}. We noted that the proper asymptotic regime  would be experimentally difficult to obtain since it corresponds to frequencies $\sim 10^2$ Hz.

This study can be pursued in several interesting ways : What happens when the bead is constrained to move in a fluid or on a fluid surface, and is likely to be sensitive to the waves generated by itself ? Recent studies have shown the extraordinary behaviors which happen in such composite systems of a bead or droplet and an interacting fluid, when the latter has a long relaxation time \cite{CouderFortMarcheurs,Snezhko,MartinSneszkoREVIEW}. Another interesting question would be to probe the behaviour of non spherical magnets : As the asymptotic ($\om\rightarrow\infty$) is a free rolling for the sphere, how does a non spherical magnet accommodate to high excitation frequencies ? Finally, a third class of follow-ups would be to inquire the collective behaviour of several beads excited together by the magnetic stirrer. Such systems may display emergent properties, typical of dissipative-active systems, where a continuous flux of energy drives assemblies of particles far from equilibrium into a unexpected stationary and complex dynamical regimes.

\section{Acknowledgements}

We thank {numerous} students who worked during their internship on this experiment or on variants of it : Elodie Adam, Vanessa Bach, J\'er\'emie Geoffre, Vincent Hardel, Alexandre Ohier, Yona Schell and  Camille Vandersteen Mauduit-Larive.

\section{Appendices}

\subsection{Dynamical equations in  spherical coordinates}\label{app1}

For sake of completeness, we provide here the dynamical equations (Eq. \myref{nseconde})  for the magnet axis $\bm n$ in the spherical coordinates defined by $\bm n=\cos\te\bm e_r+\sin\te(\sin\phi \bm e_\al-\cos\phi\bm e_z)$ :
\begin{align}
  \mcal{V}_r&=\dot{r}-\eps\left(\dot\te\cos\phi+\dot\al\sin^2\te\sin\phi\cos\phi\right.\nonumber\\
  &\left.\ \ \ \ -\dot\phi\sin\te\cos\te\sin\phi+\dot\psi\sin\te\sin\phi\right),\label{Vr}\\
  \mcal{V}_\al&=r\dot\al +\eps( \dot\phi\sin^2\te+\dot\al\sin\te\cos\te\cos\phi+\dot\psi\cos\te).\label{Valpha}
\end{align}
\begin{multline}
  \frac{d}{dt}\left[\intvide\dot\te+\dot\al\sin\phi\right]=\dot\phi^2\sin\te\cos\te+\dot\al^2\sin\te\cos\te\sin^2\phi\\+\dot\al\dot\phi\cos(2\te)\cos\phi
  +\frac{5\kappa}{2}\frac{\pa \bm n}{\pa\te}\cdot\bm B+\frac{5}{2\eps}\mcal{V}_r\cos\phi,
\end{multline}
\begin{multline}
  \frac{d}{dt}\left[\dot\phi\sin^2\te+\dot\al\sin\te\cos\te\cos\phi\right]=\dot{\al}^2\sin^2\te\sin\phi\cos\phi\\
  +\dot\al\dot\te\cos\phi-\dot\al\dot\phi\sin\te\cos\te\sin\phi\\+\frac{5\kappa}{2}\frac{\pa\bm n}{\pa\phi}\cdot\bm B-\frac{5\kappa}{2\eps}[\mcal{V}_r\sin\te\cos\te\sin\phi+\mcal{V}_\al\sin^2\te].
\end{multline}
\begin{align}
  \ddot\psi&=\frac{5}{2\eps}[\mcal{V}_r\sin\te\sin\phi-\mcal{V}_\al\cos\te].
\end{align}

\subsection{Why is rapid corotation never observed ?}\label{app:why}

In the preceding analysis, we found only a counter-rotating regime (i.e. $\om\dot\al<0$) because we {have assumed} $\phi=-\Phi+\chi=\om t-\alpha+\chi$. Another possibility to have a nonzero radial magnetic force resisting the centrifugal force would have been to write $\phi=\Phi+\chi$. In this case, we would find a corotative regime, with $\dot\al=\eps\om/(r+\eps)$ and $\dot\phi=-\om r/(r+\eps)$, and $\chi=\pi$ because the stability criterion assumed above is obviously still valid. To understand why this corotative regime is  observed neither in the experiments nor in the simulations, we assume for sake of simplicity that the driving frequency is so high that we can disregard the friction term $\lan \mcal{V}_r\ran$ in \myref{eq21}. We also model the magnetic field experienced by the bead by the orthoradial structure $\bm{B}\simeq \hat{B}(r)[\sin(\Phi) \bm e_\al+\cos(\Phi)\bm e_z]$ where $\hat{B}(r)$ is the typical field amplitude along the trajectory at mean radius $r$. Notice that the trigonometric factors in this formula are dictated by the geometrical structure of the field, see fig. \ref{fig:portrait}. The mean magnetic force resisting the centrifugal force is $F_{\rm mag}=\kappa\eps^2\pa_r\lan B_\al\sin\phi-B_z\cos\phi\ran$. With the counter-rotative Ansatz $\phi=-\Phi+\pi$, we have $F_{\rm mag}=\kappa\eps^2\pa_r\hat{B}(r)$, which is negative (as required) for typical values of $r$ larger than the value where $\hat{B}(r)$ is maximum.
With the corotative Ansatz $\phi=\Phi+\pi$, we would have $F_{\rm mag}=-\kappa\eps^2\pa_r\hat{B}(r)\lan \cos(2\Phi+\pi)\ran=0$. In fact, this value is not strictly zero, since we made approximations concerning the structure of the field. However, it is small and therefore precludes the stabilization of a corotative motion.

\bibliographystyle{unsrt}
\bibliography{Refs.bib}
\end{document}